\documentclass[10pt,conference,a4paper]{IEEEtran}
\usepackage{hyperref,times,amsmath,epsfig,graphicx,zed-csp,alltt,paralist}
\usepackage{fancyhdr,url}
\makeatletter
\def\url@leostyle{%
  \@ifundefined{selectfont}{\def\UrlFont{\sf}}{\def\UrlFont{\small\ttfamily}}}
\makeatother
\newcommand{\go}[1]{\stackrel{#1}{\longrightarrow}}
\renewcommand{\seq}{\mbox{~\large{;}}~}
\newcommand{\sq}{\mbox{~;~}}

\newcommand{\cpair}{\div}

\newcommand{\close}[1]{[~#1~]}

\newtheorem{theorem}{Theorem}
\newtheorem{lemma}{Lemma}

\title{Deriving Relationship Between Semantic Models - An Approach for cCSP}
\author{%
{Shamim H. Ripon{\small $~^{1}$}, Michael Butler{\small $~^{2}$} \thanks{S. Ripon is the corresponding author}  }%
\vspace{1.6mm}\\
\fontsize{10}{10}\selectfont\itshape
$^{1}$\,Department of Computing Science, University of Glasgow, UK\\

\fontsize{9}{9}\selectfont\ttfamily\upshape
\\
\fontsize{10}{10}\selectfont\rmfamily\itshape
$^{2}$\,School of Electronics and Computer Science, University of Southampton, UK\\

\fontsize{9}{9}\selectfont\ttfamily\upshape
}

\begin{document}
\maketitle
\renewcommand{\headrulewidth}{0.0pt}
\thispagestyle{fancy}

\setcounter{page}{47}
\pagestyle{fancy}
\rhead{\emph{(IJCSIS) International Journal of Computer Science and Information Security,\\
Vol. 7, No. 1, 2010}}
\rfoot{{\textsf{\url{http://sites.google.com/site/ijcsis/}\hspace*{.5in} \\
ISSN 1947-5500}\hspace*{1.5in}}}

\begin{abstract}
Formal semantics offers a complete and rigorous definition of a language. It is important to define different semantic models for a language and different models serve different purposes. Building equivalence between different semantic models of a language strengthen its formal foundation. This paper shows the derivation of denotational semantics from operational semantics of the language cCSP. The aim is to show the correspondence between operational and trace semantics. We extract traces from operational rules and use induction over traces to show the correspondence between the two semantics of cCSP.
\end{abstract}

\section{Introduction}\label{sec:intro}
A formal semantics offers a complete, and rigorous definition of a language. Operational and denotational semantics are two well-known methods of assigning meaning to programming languages and both semantics
are necessary for a complete description of the language.
Denotational semantics associates an element of a semantic domain
to each expression in the language and the semantic is
compositional. Traces are one of the ways to define denotational
semantics. A trace gives the global picture of the behaviour. The
common way of defining operational semantics is to provide state
transition systems for the language, where the transition system
models the computation steps of expressions in the language and
allows the formal analysis of the language.

\emph{Compensating CSP} (cSCP)~\cite{csp25} is a language defined to model long running business transactions within the framework of Hoare's CSP~\cite{Hoare:CSP} process algebra. Business transactions need to deal with faults that can arise at any stage of the transactions. \emph{Compensation} is defined in \cite{gray:tranproc} as an action taken to recover from error in business transactions or cope with a change of plan. cCSP provides constructs for orchestration of compensations to model business transactions. With the introduction of the language, both traces~\cite{csp25} and operational~\cite{cCSP05} semantics have been defined. Both semantics have valuable non-overlapping application and we want to use them both. The key question is \emph{"How they are related?"}.

This paper draws the correspondence of two different semantic
representation of a language which strengthen the formal
foundation of the language. In particular, the aim is to
accomplish the unification between operational and denotational
approach of cCSP. The unification is based on the approach where
we use the transition rules from operational semantics to derive
the traces and then show that these derived traces correspond to
the original traces by using induction over the derived traces.
Completing the derivation means that any of the presentations can
be accepted as a primary definition of the meaning of the language
and each of the definitions can even safely and consistently be
used at different times and for different purposes.

The reset of the paper is organised as follows. A brief overview of cCSP along with an example is given in Section~\ref{sec:ccsp}. The trace and the operational semantics of cCSP are outlined in Section~\ref{sec:semantics}. We describe the how we define and prove a relationship between the semantic models in Section~\ref{sec:relation}. We define theorems and supporting lemmas to establish the relationship for both standard and compensable processes. We outline some lessons from the experiment and then summarise some related work in Section~\ref{sec:lesson} and Section~\ref{sec:relwork} respectively. We draw our conclusion in Section~\ref{sec:concl}.

\section{Compensating CSP}\label{sec:ccsp}

\begin{figure}[!htb]
\centerline{\psfig{figure=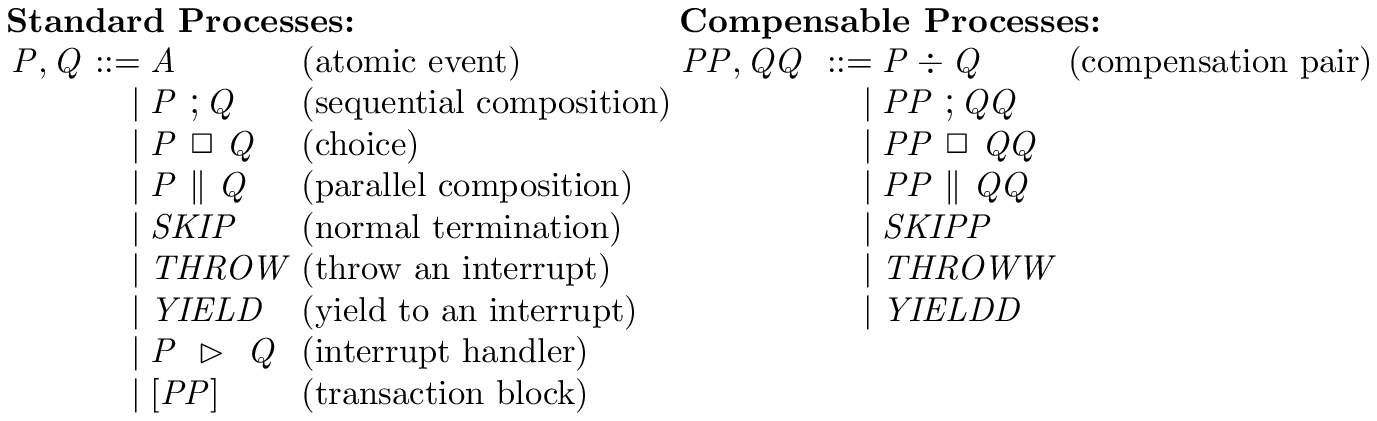,width=88.54mm} }
\caption{cCSP syntax}
\label{fig:syntax}
\end{figure}

The introduction of the cCSP language was inspired by two ideas: transaction processing features, and process algebra. Like standard CSP, processes in cCSP are modelled in terms of the atomic events they can engage in. The language provides operators that support sequencing, choice, parallel composition of processes. In order to support failed transaction, compensation operators are introduced. The processes are categorised into \emph{standard}, and \emph{compensable} processes. A standard process does not have any compensation, but compensation is part of a compensable process that is used to compensate a failed transaction. We use notations, such as, $P,Q,..$ to identify standard processes, and $PP,QQ,..$ to identify compensable processes. A subset of the original cCSP is considered in this paper, which includes most of the operators, is summarised in Fig.~\ref{fig:syntax}.

The basic unit of the standard processes is an atomic event ($A$). The other operators are the sequential~($P\sq Q$), and the parallel composition ($P\parallel Q$), the choice operator ($P\extchoice Q$), the interrupt handler ($P\rhd Q$), the empty process $SKIP$, raising an interrupt $THROW$, and yielding to an interrupt $YIELD$. A process that is ready to terminate is also willing to yield to an interrupt. In a parallel composition, throwing an interrupt by one process synchronises with yielding in another process. Yield points are inserted in a process through $YIELD$. For example, ($P\sq YIELD\sq Q$) is willing to yield to an interrupt in between the execution of $P$, and $Q$. The basic way of constructing a compensable process is through a compensation pair ($P\cpair Q$), which is constructed from two standard processes, where $P$ is called the \emph{forward} behaviour that executes during normal execution, and $Q$ is called the associated compensation that is designed to compensate the effect of $P$ when needed. The sequential composition of compensable processes is defined in such a way that the compensations of the completed tasks will be accumulated in reverse to the order of their original composition, whereas compensations from the compensable parallel processes will be placed in parallel. In this paper, we define only the asynchronous composition of processes, where processes interleave with each other during normal execution, and synchronise during termination. By enclosing a compensable process $PP$ inside a transaction block $\close{PP}$, we get a complete transaction and the transaction block itself is a standard process. Successful completion of $PP$ represents successful completion of the block. But, when the forward behaviour of $PP$ throws an interrupt, the compensations are executed inside the block, and the interrupt is not observable from outside of the block. $SKIPP, THROWW$, and $YIELDD$ are the compensable counterpart of the corresponding standard processes and they are defined as follows:
\begin{eqnarray*}
SKIPP &=& SKIP\cpair SKIP,\\
YIELDD &=& YIELD\cpair SKIP\\
THROWW &=& THROW\cpair SKIP
\end{eqnarray*}

\begin{figure}[htb]
\centering
\includegraphics[width=88 mm]{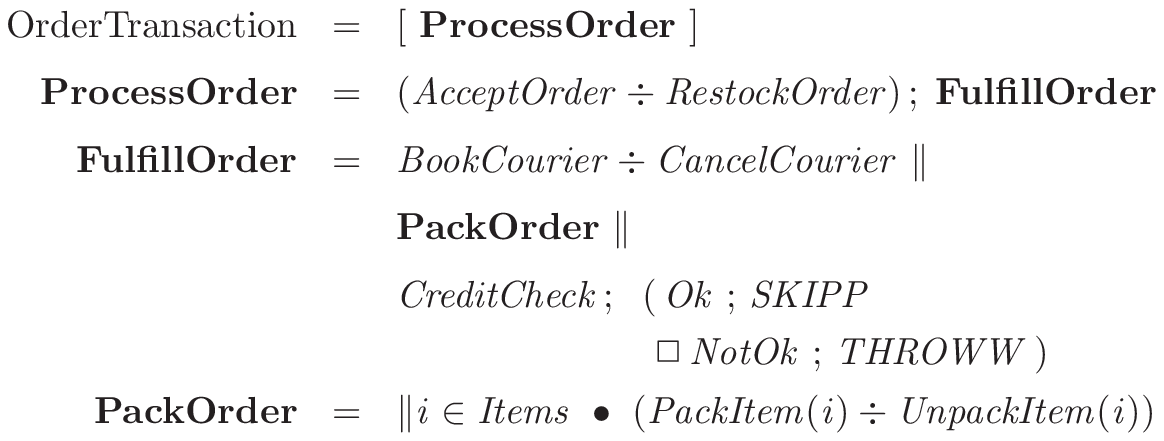}
\caption{Warehouse order processing}
\label{fig:example}
\end{figure}

To illustrate the use of cCSP, we present an example of a transaction for processing customer orders in a warehouse in~Fig.\ref{fig:example}. The first step in the transaction is a compensation pair. The primary action of this pair is to accept the order and deduct the order quantity from the inventory database. The compensation action  simply  adds the order quantity back to the total in the inventory database. After an order is received from a customer, the  order is packed for shipment, and a courier is booked to deliver the goods to the customer. The $\textbf{PackOrder}$ process packs each of the items in the order in parallel. Each $\mathit{PackItem}$ activity can be compensated by a corresponding $\mathit{UnpackItem}$. Simultaneously with the  packing of the order,  a credit check is performed on the customer. The credit check is performed in parallel because it normally succeeds, and in this normal case the company does not wish to delay the order unnecessarily. In the case that a credit check fails, an interrupt is thrown causing the transaction to stop its execution, with the courier possibly having been booked and possibly some of the items having being packed. In case of failure, the semantics of the transaction block will ensure that the appropriate compensation activities will be invoked for those activities
that already did take place.

\section{Semantic Models}\label{sec:semantics}

This section briefly outlines the trace and the operational semantics of cCSP.
\subsection{Trace Semantics}

A trace of a process records the history of behaviour up to some
point. We show the operators on traces which are then
lifted to operators on set of traces. Traces considered for cCSP
are non-empty sets.

\begin{figure}[!htb]
\centering
\framebox{
\includegraphics[width=86 mm]{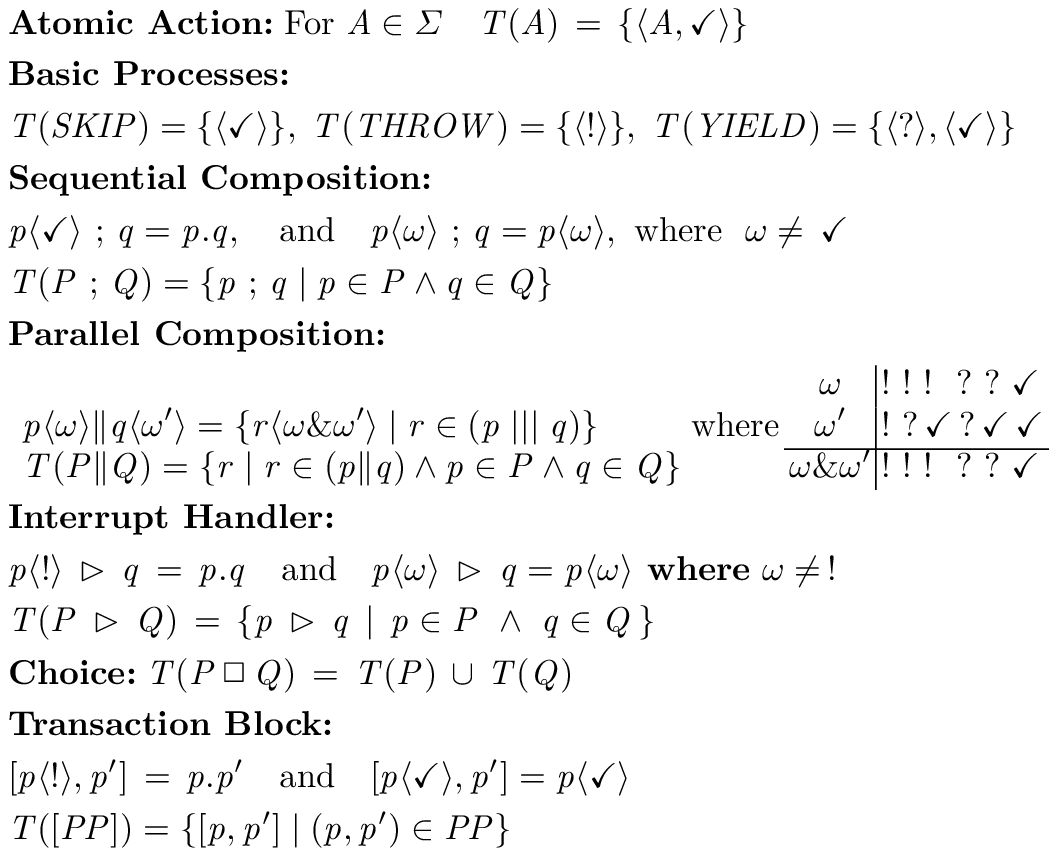}}
\caption{Trace semantics of standard processes}
\label{fig:stdtrace}
\end{figure}

The trace of a standard process is of the form $s\trace{\omega}$
where $s\in\Sigma^{*}$ ($\Sigma$ is alphabet of normal events) and
$\omega\in\Omega$ ($\Omega=\{\tick,~!,~?\}$), which means all
traces end with any of the events in $\Omega$, which is called a
terminal event. The terminal events represent the termination of a
process. Successful termination is shown by a $\tick$. Termination
by either throwing or yielding an interrupt is shown by $!$ or $?$
respectively. In sequential composition $(p\sq q)$, the
concatenated observable traces $p$ and $q$, only when $p$
terminates successfully,(ends with $\tick$), otherwise the trace
is only $p$. The traces of two parallel processes are
$p\trace{\omega}\Vert q\trace{\omega'}$ which corresponds to the
set ($p\interleave q$), the possible interleaving of traces of
both processes and followed by $\omega\&\omega'$, the
synchronisation of $\omega$ and $\omega'$. The trace semantics of standard processes are shown in Fig.~\ref{fig:stdtrace}.

\begin{figure}[!htb]
\centering
\framebox{
\includegraphics[width=86 mm]{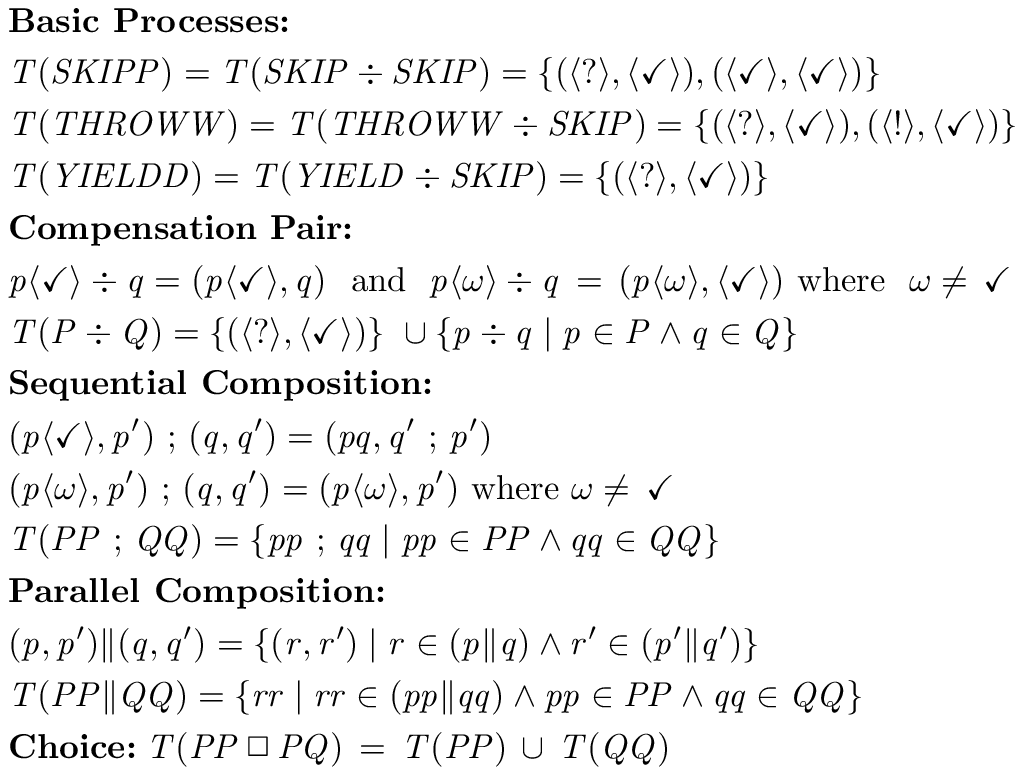}}
\caption{Trace semantics of compensable processes}
\label{fig:comptrace}
\end{figure}

Compensable processes are comprised of \emph{forward} and
\emph{compensation} behaviour. The traces of compensable processes
are of pair of traces of the form
$(s\trace{\omega},s'\trace{\omega'})$, where $s\trace{\omega}$ is
the forward behaviour and $s'\trace{\omega'}$ is the compensation
behaviour. In sequential composition, the forward traces correspond
to the original forward behaviour and followed by the traces
of the compensation. Traces of parallel composition are defined as
the interleaving of forward traced and then follows the
interleaving of compensation. The traces of a compensation pair
are the traces of both of the processes of the pair when the
forward process ($P$) terminate with a $\trace{\tick}$, otherwise
the traces of the pair are the traces of the forward process
followed by only a $\trace{\tick}$. The traces of a transaction block are only the traces of compensable processes inside the block when the
process terminates with a $\trace{\tick}$, otherwise when the
forward process inside the block terminates with a $\trace{!}$ the
traces of the block are the traces of the forward process followed
by the traces of the compensation. Fig.~\ref{fig:comptrace} outlines the traces of compensable processes.

The following healthiness conditions declare that processes consist of some terminating or interrupting behaviour which ensures that the traces of processes are non-empty:
\begin{itemize}
\item $p\trace{\tick}\in T(P)$ ~or~ $p\trace{~!~}\in T(P)$, for some $p$
\item $(p\trace{\tick},p')\in T(PP)$ or $(p\trace{~!~},p')\in T(PP),$ for some $p,p'$
\end{itemize}

\subsection{Operational Semantics}

By using labelled transition systems \cite{Plotkin:OS}, the
operational semantics specifies the relation between states of a
program. Two types of transitions are define to present the
transition relation of process terms: normal and terminal. A normal transition is defined by a normal event ($a\in\Sigma$) and a terminal transition is defined by a terminal event ($\omega\in\Omega$) .

For a standard process, a normal transition makes the transition of a process term from one state to its another state ($P$ to $P'$). The
terminal transition, on the other hand terminates a
standard process to a null process (0):
\begin{eqnarray*}
    P \go{a}P', &&
    P \go{\omega}0
\end{eqnarray*}

\begin{figure}[!htb]
\centering
\framebox{
\includegraphics[width=86 mm]{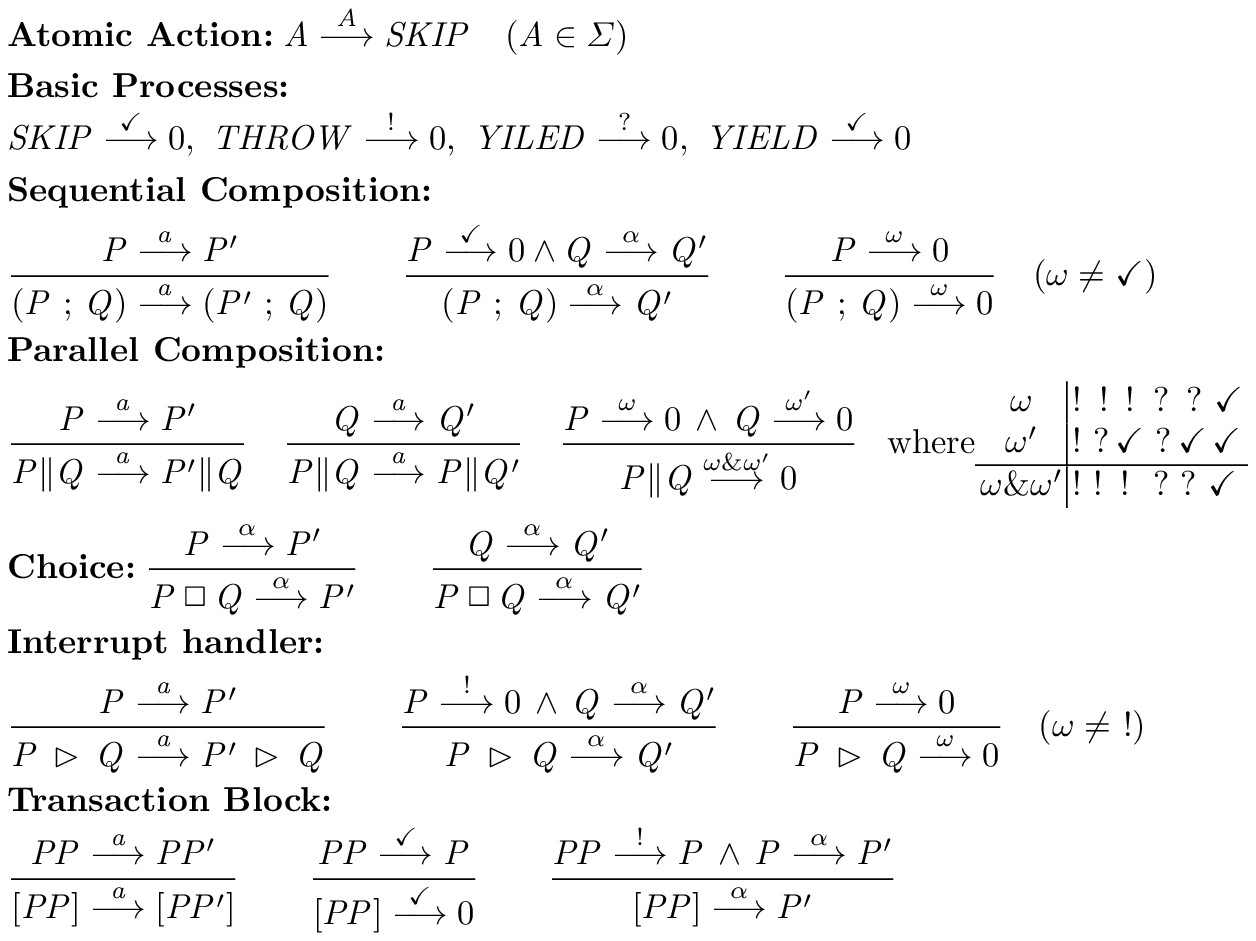}}
\caption{Operational semantics of standard processes}
\label{fig:std-os}
\end{figure}

In sequential composition ($P\sq Q$), the process $Q$ can start only when the process $P$ terminates successfully (with $\tick$). If $P$ terminates with $!$ or $?$ the process $Q$ will not start. In parallel composition each process can evolve independently and processes synchronise only on terminal events. The transition rules for standard processes are outlined in Fig.~\ref{fig:std-os}.

For compensable processes, the normal transitions are same as standard processes. However, the terminal events terminate the forward behaviour of compensable processes, additionally, the compensation are stored for future reference.
\begin{eqnarray*}
    PP \go{a}PP', &&
    PP \go{\omega} P \quad \mbox{($P$ is the compensation)}
\end{eqnarray*}

\begin{figure}[!htb]
\centering
\framebox{
\includegraphics[width=86 mm]{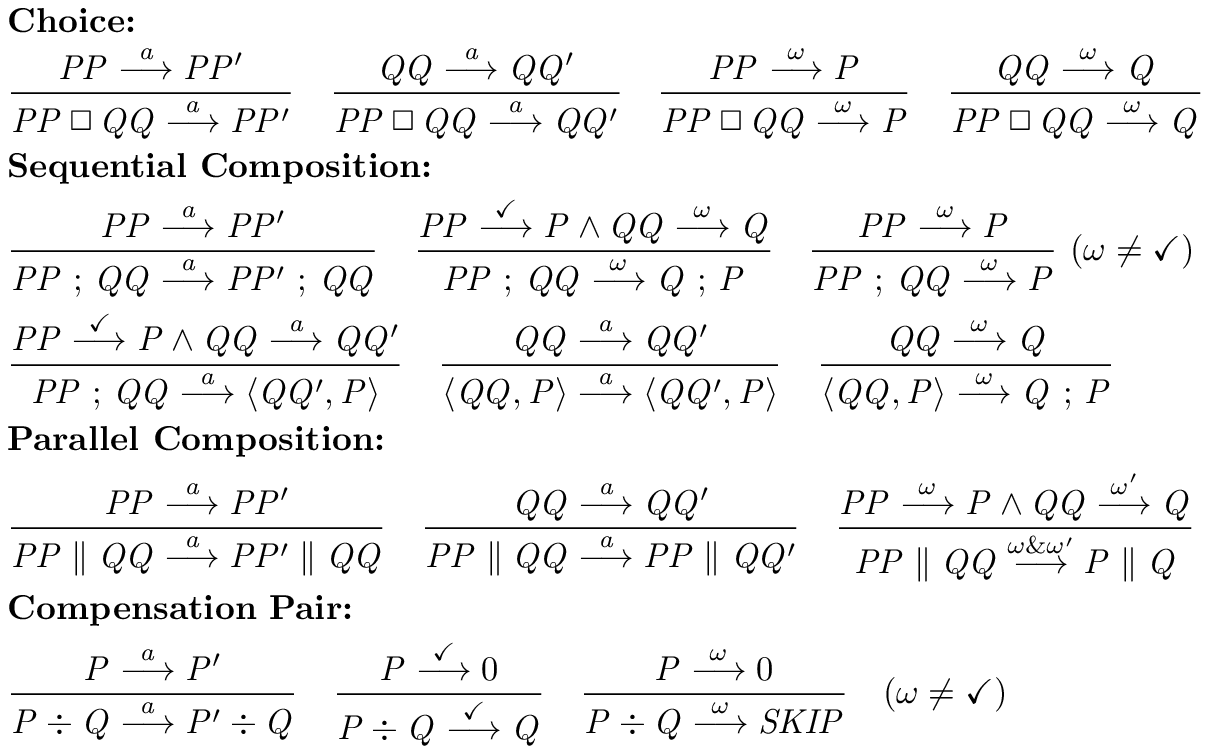}}
\caption{Operational semantics of compensable processes}
\label{fig:comp-os}
\end{figure}

In sequential composition ($PP\sq QQ$), when $PP$ terminates, its compensation ($P$) is stored and $QQ$ starts to execute. In this scenario, we get an auxiliary construct ($\trace{QQ,P}$) where the processes have no particular operational relation between them. After termination of the process $QQ$, its compensation ($Q$) is accumulated in front of $P$ i.e., ($Q\sq P$). In the parallel composition, the main difference with the standard processes is that after termination of the forward behaviour the compensations are accumulated in parallel. The transition rules of compensable processes are summarised in Fig.~\ref{fig:comp-os}.

A non-terminal event changes the state of the process inside the
block. Successful completion of the forward process inside the
block means completion of the whole block, but throwing a
interrupt by the compensable process inside the block results the
compensation to run.  In compensation pair, after successful
completion of the forward behaviour the compensation will be
stored for future use, however, unsuccessful termination, i.e,
terminates by $!$ or $?$ results an empty compensation (Fig.~\ref{fig:std-os}).

\section{Relating Semantic Models}\label{sec:relation}

\begin{figure}[!htb]
\centering
\includegraphics[width=86 mm]{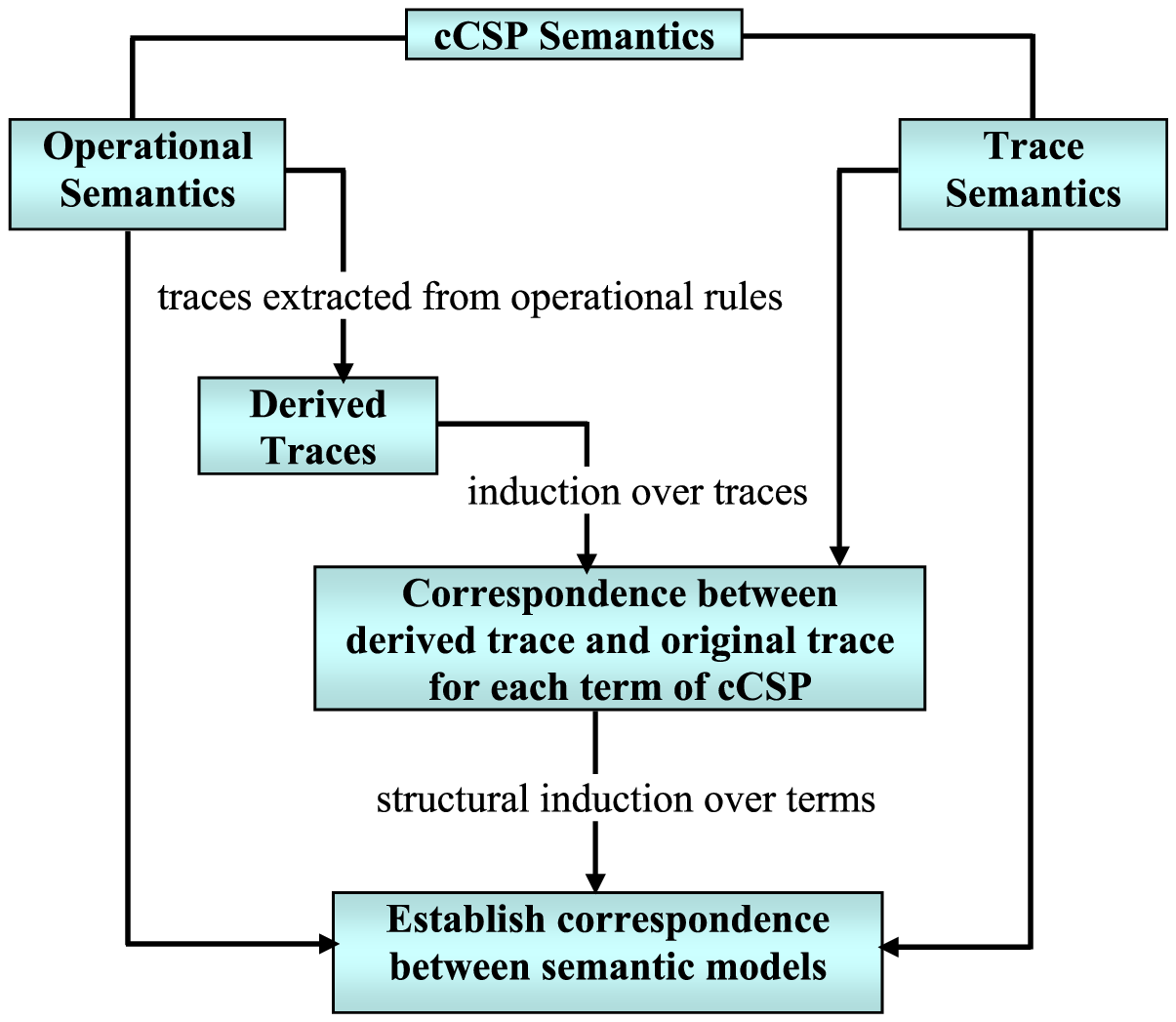}
\caption{Steps to derive relationship between semantic models}
\label{fig:proof-steps}
\end{figure}

In this section we describe the steps to derive a relationship between the two semantic models of cCSP. We follow a systematic approach to derive the relationship where traces are first extracted from the transition rules and prove that the extracted traces correspond to the original trace definition. The steps of deriving the semantic relation are shown in Fig.~\ref{fig:proof-steps}.

The operational semantics leads to lifted transition relations
labelled by sequences of events. This is defined
recursively. For a standard  process $P$,
\begin{eqnarray*}
    P\go{\trace{\omega}}Q &~=~& P\go{\omega}Q\\
    P\go{\trace{a}t}Q &~=~& \exists P'\cdot
    P\go{a}P'~\land~P'\go{t}Q
\end{eqnarray*}
The derived traces of a standard process $P$ is defined as $DT(P)$. Let $t\in DT(P)$, then we get the following definition,
\begin{eqnarray}
    t\in DT(P) &=& P\go{t}0\label{eq:dtp}
\end{eqnarray}

Compensable processes have both forward and compensation
behaviour. A compensable process is defined as a pair of traces. Hence, it is required
to extract traces from both forward and compensation behaviour.
The forward behaviour of a compensable process $PP$ is defined as
follows:
\begin{eqnarray*}
    PP\go{t}R \quad(\mbox{$t$ ends with $\omega$})
\end{eqnarray*}
where $t$ is the trace of the forward behaviour.  $R$ is the attached
compensation. The behaviour of compensation is similar to standard
processes and by reusing that we get the following definition:
\begin{eqnarray*}
    PP\go{(t,t')}0 &~=~& \exists R\cdot PP\go{t}R~ \land~R\go{t'}0
\end{eqnarray*}
where $t'$ is the trace of the compensation. For a compensable process $PP$, the derived traces $DT(PP)$ is defined as follows:
\begin{eqnarray*}
    (t,t')\in DT(PP) &~=~& PP\go{(t,t')}0
\end{eqnarray*}

By using the definition of derived traces and the original traces we state the following theorem to define the relationship between the semantic models,
\noindent\begin{theorem}\label{th1}
For any standard process term $P$, where $P\neq 0$
\begin{eqnarray*}DT(P)~~=~~T(P)\end{eqnarray*}
For any compensable process terms $PP$, where $PP\neq 0$ and does
not contain the term $\trace{PP,P}$,
\begin{eqnarray*}
  DT(PP) &~=~& T(PP)
\end{eqnarray*}
\end{theorem}

Traces are extracted for each term of the language, and its correspondence is shown with the corresponding traces in the trace semantics. Assume $P$ and $Q$ are standard process terms, then for all the operators, we prove that
 \begin{eqnarray}
   t\in DT(P\otimes Q)&~~=~~& t\in T(P\otimes Q)\label{eq:genstd}
 \end{eqnarray}
For each such operator $\otimes$, the proof is performed by
induction over traces. In the proof we assume that,
$DT(P)~=~ T(P)$ and $DT(Q) ~=~ T(Q)$.

We follow similar style for compensable processes. Assuming $DT(PP)~=~ T(PP)$ and $DT(QQ) ~=~ T(QQ)$ we show that,
\begin{eqnarray}
 (t,t')\in DT(PP\otimes QQ)~~ &=& (t,t')\in T(PP\otimes QQ)
 \label{eq:gencomp}
\end{eqnarray}
In the following sections we outline the proof steps showing the correspondence in (\ref{eq:genstd}) and (\ref{eq:gencomp}) for both standard and compensable process terms.

\subsection{Standard Processes}

\noindent\textbf{Sequential Composition:} By using (\ref{eq:genstd}) the relationship between the semantic models is derived by showing that,
$$t\in DT(P\sq Q)~~=~~t\in T(P\sq Q)$$
From (\ref{eq:dtp}) we  get the derived traces of the sequential composition,
\begin{eqnarray*}
t\in DT(P\sq Q) &=& (P\sq Q)\go{t}0
\end{eqnarray*}
We also expand the definition of trace semantics as follows:
\begin{align*}
    t &\in~ T(P\seq Q)&&\\
      &=~\exists p,q\cdot t=(p\sq q)~~\land~~
        p\in T(P)~~\land~~q\in T(Q) &&\\
      &=~\exists p,q\cdot t=(p\sq q)~~\land~~
        p\in DT(P)~~\land~~q\in DT(Q) &&\\
      &=~\exists p,q\cdot t=(p\sq q)~~\land~~
      P\go{p}0~~\land~~Q\go{q}0&&
\end{align*}
Finally, from the above definitions of traces, the following lemma is formulated for the sequential composition of standard processes:

\noindent\begin{lemma}\label{lema:stseq}
\begin{eqnarray*}
(P\sq Q) \go{t}0 ~=~\exists p,q\cdot t=(p\sq q)~\land~
P\go{p}0~\land~Q\go{q}0
\end{eqnarray*}
\end{lemma}
The lemma is proved by applying induction over the trace $t$, where $t=\trace{\omega}$ is considered as the base case, and $t=\trace{a}t$ is considered as the inductive case. To support the proof of the lemma, two equations are derived from the transition rules. These derived equations are based on the event by which the transition rules are defined:
\begin{eqnarray*}
(P\sq Q)\go{\omega}0 &~=~& P\go{\tick}0~\land~Q\go{\omega}0\\
	&\vee & P\go{\omega}0~\land~\omega\neq\tick \label{eq3}
\\
(P\sq Q)\go{a}R &~=~&\exists P'\cdot P\go{a}P'~\land~R=(P'\sq Q)\\
				&\vee & P\go{\tick}0~\land~Q\go{a}R\label{eq4}	 
\end{eqnarray*}

\begin{proof}~\\
\textbf{Basic step}:\; $t=\trace{\omega}$\\
$\begin{array}{rcl}
    (P\sq Q)\go{\trace{\omega}}0 &=&
    (P\sq Q)\go{\omega}0
\end{array}$\\
``From transition rules sequential composition''
\begin{eqnarray}
    &=& ~~~~P\go{\tick}0~~\wedge~~ Q\go{\omega}0\label{steqn1}\\
    &\vee&  ~P\go{\omega}0~~\wedge~~\omega \neq\tick\label{steqn2}
\end{eqnarray}
From (\ref{steqn1})
\begin{eqnarray*}
    &&   P \go{\tick}0~~\wedge~~ Q \go{\omega}0\\
 &=& \exists p,q\cdot~p=\trace{\tick}~\wedge~q=\trace{\omega}~
        \wedge~\trace{\omega}=(p\sq q) \\
 &\wedge& P\go{p}0~\wedge~ Q \go{q}0\\
    &=&  \exists p,q\cdot\trace{\omega}= (p\sq q)~~\wedge~~p=\trace{\tick}\\
   &\wedge&~P\go{p}0~~\wedge~~ Q\go{q}0
\end{eqnarray*}
From (\ref{steqn2})
\begin{eqnarray*}
    &&   P\go{\omega}0~~\wedge~~\omega\neq\tick\\
    &=&   \exists p,q\cdot p=\trace{\omega}~~\wedge~~ \omega\neq\tick
        ~~\wedge~~\trace{\omega}=(p\sq q)\\
    &\wedge& P\go{p}0~~\wedge~~ Q\go{q}0\\
    &=&  \exists p,q\cdot \trace{\omega}=(p\sq q)~~\wedge~~ p\neq\trace{\tick}\\
   &\wedge& P\go{p}0~~\wedge~~ Q\go{q}0
\end{eqnarray*}
Therefore, for $t=\trace{\omega}$ from (\ref{steqn1}) and (\ref{steqn2})
\begin{eqnarray*}
    &&   \exists p,q\cdot\trace{\omega}= (p\sq q)
        ~~\wedge~~p=\trace{\tick}\\
    &&\wedge~ P\go{p}0~~\wedge~~ Q\go{q}0\\
    &\vee& \exists p,q\cdot \trace{\omega}=(p\sq q)
        ~~\wedge~~ p\neq\trace{\tick}\\
  &&\wedge~ P\go{p}0~~\wedge~~ Q\go{q}0\\
   &=&   \exists p,q\cdot \trace{\omega}=(p\sq q)
        ~~\wedge~~ P\go{p}0~~\wedge~~ Q\go{q}0
\end{eqnarray*}
\textbf{Inductive step:} $t=\trace{a}t$\\
$\begin{array}{rcl}
    P\seq Q\go{\trace{a}t}0 &=&
    \exists R\cdot~(P\sq Q)\go{a}R~~\wedge~~R\go{t}0
\end{array}$\\
\\``From operational rules''
\begin{eqnarray}
&=& ~~~~\exists P'\cdot~P\go{a}P'
    ~~\wedge~~(P'\sq Q)\go{t}0 \label{steqn3}\\
&&  \vee~\exists Q'\cdot~P\go{\tick}0
    ~~\wedge~~Q\go{a}Q'~~\wedge~~Q'\go{t}0\label{steqn4}
\end{eqnarray}
From ~(\ref{steqn3})
\begin{eqnarray*}
&& \exists P'\cdot~P\go{a}P'~~\wedge~~(P'\sq Q)\go{t}0 \\
&=& \mbox{``Inductive hypothesis"}\\
&&  \exists P'\cdot~P\go{a}P'~~\wedge~~
    \exists p',q\cdot~t=(p'\sq q)\\
 &\wedge& P'\go{p'}0~~\wedge~~ Q\go{q}0\\
&=&  \mbox{``Combining existential quantifications"}\\
&&   \exists p',q\cdot~t=(p'\sq q)~~\wedge~~ P\go{\trace{a}p'}0
    ~~\wedge~~ Q\go{q}0\\
&=&  \mbox{``Using trace rule~~}
    \trace{a}t=\trace{a}(p'\sq q)=(\trace{a}p')\sq q~\mbox{"}\\
&&   \exists p',q\cdot~\trace{a}t=(\trace{a}p'\sq q)~~\wedge~~
    P\go{\trace{a}p'}0~~\wedge~~ Q\go{q}0\\
&=&  \exists p,q\cdot p=\trace{a}p' ~~\wedge~~\trace{a}t=(p\sq q)\\
&\wedge& P\go{p}0~~\wedge~~ Q\go{q}0
\end{eqnarray*}
From ~(\ref{steqn4})
\begin{eqnarray*}
  &&  \exists Q'\cdot~ P\go{\tick}0
    ~~\wedge~~Q\go{a}Q'~~\wedge~~Q'\go{t}0\\
  &=& P\go{\tick}0~~\wedge~~ Q\go{\trace{a}t}0\\
  &=& \exists p,q\cdot~p=\trace{\tick}~~\wedge~~
      q=\trace{a}t~~\wedge~~\trace{a}t=(p\sq q)\\
  &\wedge& P\go{p}0~~\wedge~~ Q\go{q}0\\
  &=& \exists p,q\cdot~\trace{a}t=(p\sq q)~~\wedge~~p=\trace{\tick}\\
   &\wedge& P\go{p}0~~\wedge~~ Q\go{q}0
\end{eqnarray*}
Therefore for $t=\trace{a}t$, from (\ref{steqn3}) $\vee$
(\ref{steqn4})
\begin{flalign*}
   &~\exists p,q\cdot p=\trace{a}p'~~\wedge~~\trace{a}t=(p\sq q)
        ~~\wedge~~ P\go{p}0~~\wedge~~ Q\go{q}0\\
   \vee &~ \exists p,q\cdot~p=\trace{\tick}~~\wedge~~
        \trace{a}t=(p\sq q) ~~\wedge~~  P\go{p}0~~\wedge~~ Q\go{q}0\\
   =&~ \mbox{``Combining existential quantifications"}\\
   &~\exists p,q\cdot~(p=\trace{\tick}~\vee~p=\trace{a}p')
        ~~\wedge~~\trace{a}t=(p\sq q)\\
   &\wedge~ P\go{p}0~~\wedge~~Q\go{q}0\\
   =&~ \exists p,q\cdot \trace{a}t=(p\sq q)
        ~~\wedge~~ P\go{p}0~~\wedge~~ Q\go{q}0
\end{flalign*}
\end{proof}
This completes the proof of the lemma. We follow the same approach to prove other lemmas in the rest of the paper.

\noindent\textbf{Parallel Composition:}\; The parallel composition of two processes is defined to be the interleaving of their observable events followed by the synchronisation of their terminal events. For example, considering asynchronous actions, $A\parallel B$ can execute $A$ followed by $B$ or $B$ followed by $A$. For traces $p$ and $q$ we write $p\interleave q$ to denote the set of interleaving of $p$ and $q$ and it follows the following definition:
\begin{eqnarray*}
    \trace{}\in p\interleave q &=&  p=\trace{}~~\wedge~~q=\trace{}\\
    \trace{a}t\in p\interleave q &=& \exists p'\cdot~
        p=\trace{a}p'~~\wedge~~t\in p'\interleave q\\
        &\vee & \exists q'\cdot~q=\trace{a}q'
        ~~\wedge~~t\in p\interleave q'
\end{eqnarray*}

By following similar steps as sequential composition, we define the following lemma for parallel composition:
\begin{lemma}\label{lema4}
$$(P\parallel Q) \go{t}0 ~=~\exists
p,q\cdot~t\in (p\parallel q) ~\wedge~ P\go{p}0~\wedge~Q\go{q}0
$$
\end{lemma}
We derive two supporting equation from the transition rules of parallel composition:
\begin{eqnarray*}
 P\parallel Q\go{a}R &=& P\go{a}P'~~\wedge~~R=P'\parallel Q\\
                    &\vee& Q\go{a}Q'~~\wedge~~R=P\parallel Q'\\
 P\parallel Q\go{\omega}0 &=& P\go{\omega1}0
   ~~\wedge~~Q\go{\omega2}0~~ \wedge~~\omega\in\omega1\&\omega2
\end{eqnarray*}
\begin{proof} The proof of the base case is trivial and omitted from the presentation. The inductive case is described here:
\begin{eqnarray*}
    &&  (P\parallel Q)\go{\trace{a}t}0\\
    &=& \exists R\cdot (P\parallel Q)\go{\trace{a}}R ~~\wedge~~ R\go{t}0 \\
    &=& \mbox{``Using the operational rules"}\\
    &&  \exists P'\cdot P\go{a}P'~~\wedge~~ (P'\parallel Q)~\go{t}0\\
    &\vee& \exists Q'\cdot Q\go{a}Q'~~\wedge~~ (P\parallel Q')\go{t}0\\
    &=& \mbox{``Inductive hypothesis"}\\
    && \exists~P'\cdot~P\go{a}P'~~\wedge~~\exists p',q\cdot~t\in (p'\parallel q)\\
    &\wedge& P'\go{p'}0~~\wedge~~Q\go{q}0\\
   &\vee&\exists~Q'\cdot~Q\go{a}Q'~~\wedge~~\exists p,q'\cdot~t\in (p\parallel q')\\
    &\wedge& P\go{p}0~~\wedge~~Q'\go{q'}0\\
   &=&  \mbox{``Combining existential quantifications"}\\
   &=&  \exists p',q\cdot~t\in (p'\parallel q)~~\wedge~~
        P\go{\trace{a}p'}0~~\wedge~~Q\go{q}0\\
   &\vee&  \exists p,q'\cdot~t\in (p\parallel q')~~\wedge~~
        P\go{p}0~~\wedge~~Q\go{\trace{a}q'}0\\
   &=&  \exists p,q \cdot p=\trace{a}p'~~\wedge~~ t\in(p'\parallel q)\\
   &&\land P\go{p}0~~\wedge~~Q\go{q}0\\
   &\vee&\exists p,q \cdot q=\trace{a}q'~~\wedge~~ t\in(p\parallel q')\\
   &&\land P\go{p}0~~\wedge~~Q\go{q}0\\
   &=&  \mbox{``Combining"}\\
   &&   \exists p,q\cdot~(p=\trace{a}p'~~\wedge~~t\in(p'\parallel q)
   ~~\vee~~ q=\trace{a}q'\\
   &&\wedge~~ t\in(p\parallel q'))
       ~~\wedge~~ P\go{p}0~~\wedge~~Q\go{q}0\\
   &=& \mbox{``By the definition the interleaving of traces"}\\
   &&   \exists p,q\cdot \trace{a}t\in (p\parallel q)~~\wedge~~
        P\go{p} 0~~\wedge~~Q\go{q}0
\end{eqnarray*}
\end{proof}

\subsection{Compensable Processes}
\noindent\textbf{Sequential Composition:} For compensable processes $PP$ and $QQ$, let $(t,t')\in DT(PP\sq QQ)$ and according to trace derivation rule we get
\begin{eqnarray*}
  (t,t')\in DT(PP\sq QQ) =~ \exists R\cdot (PP\sq QQ)\go{t}R\land
   R\go{t'}0
\end{eqnarray*}
The following lemma is stated to define the relationship for the lifted forward behaviour of sequential composition of compensable processes:
\begin{lemma}\label{lema:c-seq}
\begin{eqnarray*}
(PP\sq QQ)\go{t}R &=& \exists P,Q,p,q\cdot t~=~(p\sq q)\\
&\land& PP\go{p}P\land QQ\go{q}Q\\
&\land& R~=~ COND(last(p)=\checkmark,(Q\sq P),P)\\
Where,&& COND(true,e1,e2)~=~e1\\
&& COND(false,e1,e2)~=~e2
\end{eqnarray*}
\end{lemma}
$COND$ expression is used to state that when process $PP$ terminates successfully (terminate by $\checkmark$), compensation from both $PP$ and $QQ$ are accumulated in reverse order, otherwise only compensation from $PP$ is stored. The following equations are derived from the transition rules to support the proof of the above lemma.
\begin{eqnarray*}
 (PP\sq QQ) \go{a}RR &=& PP\go{a}PP'~\land~RR=(PP'\sq QQ)\\
  &\vee& PP\go{\tick}P~\land~QQ\go{a}QQ'\\
  &\land& R=\trace{QQ',P}\\
(PP\sq QQ)\go{\omega}R &=& PP\go{\tick}P\land QQ\go{a}Q\land R=(Q\seq P)\\
&\vee&~ PP\go{\omega}P~\land~\omega\neq\tick~\land~R=P
\end{eqnarray*}

In the inductive case of
the lemma we get the following intermediate step involving the
auxiliary construct $\trace{QQ,P}$.
\begin{eqnarray}
PP\sq QQ \go{\trace{a}t}R &=&
     \exists RR\cdot~PP\seq QQ\go{a}RR~\wedge~RR\go{t}R\notag\\
&=&\exists PP'\cdot~PP\go{a}PP'~~\wedge ~~PP'\seq
    QQ\go{t}R\notag
~\\
  &\vee& \exists P,QQ'\cdot~PP\go{\tick}P~~\wedge~~QQ\go{a}QQ'\notag\\
  &\wedge& \trace{QQ',P}\go{t} R\label{eq2}
\end{eqnarray}
To deal with this we need another lemma which will support the
removal of auxiliary construct in (\ref{eq2}). This lemma
considers the situation where the forward behaviour of the
first process of sequential composition is terminated with $\tick$
and its compensation is stored and the second process of the
composition has started. Here to mention that $t$ in
(\ref{eq2}) above is a complete trace.

\begin{lemma}\label{lema3}
$$\trace{QQ,P}\go{t}R ~=~\exists Q\cdot~QQ\go{t}Q~\land~R=(Q\seq
    P)$$
\end{lemma}
The lemma is proved by induction over traces. By using this lemma, we prove Lemma~\ref{lema:c-seq} by following the similar approach of applying induction over traces.

\noindent\textbf{Parallel Composition:}Let $(t,t')\in DT(PP\parallel QQ)$ By using the trace derivation rule we get,
\begin{eqnarray*}
(t,t')\in DT(PP\parallel QQ)~=~\exists R\cdot~
(PP\parallel QQ)\go{t}R\land R\go{t'}0
\end{eqnarray*}
We then define the following lemma to establish the semantic correspondence for parallel composition of compensable processes:

\begin{lemma}\label{lema5}
~\\$\begin{array}{rl}
(PP\parallel QQ)\go{t}R &=~ \exists P,Q,p,q\cdot~t\in (p\parallel q) \\
&\land~ PP\go{p}P~\wedge~QQ\go{q}P~\wedge~R=P\parallel Q
\end{array}$
\end{lemma}
The lemma is proved by using induction over traces similar to other lemmas.

\noindent\textbf{Compensation Pair:}
A compensation pair $(P\cpair Q)$ consists of two standard processes: a standard process ($P$) and its compensation $(Q)$. The semantics of compensation pair is defined in such a way that the behaviour of the compensation $Q$ is augmented only with successfully completed forward behaviour of $P$, otherwise, the compensation is empty. For a compensation pair, we show that
\begin{eqnarray*}
(t,t')\in DT(P\cpair Q)&~=~& (t,t')\in T(P\cpair Q)
\end{eqnarray*}
To prove the semantic correspondence between the semantics model, we state the following lemma:
\begin{lemma}\label{lema:cpair}
$$(P\cpair Q)\go{(t,t')}0~=~\exists p,q\cdot (t,t')=(p\cpair Q)\land
P\go{p}0\land Q\go{q}0 $$
\end{lemma}
The lemma is proved by induction as previous lemmas. To support the inductive proof, the following two equations are derived from the transitions rules shown earlier,
\begin{eqnarray*}
(P\cpair Q)\go{\omega}R &~=~& P\go{\tick}0~~\land~~R=Q\\
 &\vee& P\go{\omega}0~~\land~~\omega\neq\tick~\land~ R=SKIP \\
(P\cpair Q)\go{a}RR &~=~& P\go{a}P'~~\land~~RR=P'\cpair Q
\end{eqnarray*}
Unlike the lemmas defined earlier for compensable
processes, Lemma~\ref{lema:cpair} includes the traces of both forward and compensation behaviour. The following trace rules for the compensation pair are used in the proof of the lamma:
\begin{eqnarray*}
    \mbox{when}~~ p=p'\trace{\tick}
    (t,t')&=&(p'\trace{\tick}\cpair q)~=~ (p,q)\\
\mbox{when}~p = p'\trace{\omega}\land
\omega\neq\tick
    (t,t') &=& (p'\trace{\omega}\div q)~=~(p,\trace{\tick})
\end{eqnarray*}

\noindent\textbf{Transaction Block:} Transaction block is a standard process. We let $t\in DT([PP])$ and by following the trace derivation rule we get
\begin{eqnarray*}
t\in DT([PP]) &=& [PP]\go{t}0
\end{eqnarray*}
The semantic correspondence is then derived by proving the following lemma:
\begin{lemma}\label{lema-block1}
~\\$[PP]\go{t}0
    =\exists p,p'\cdot t=[p,p']~~\wedge~~PP\go{p,p'}0        $
\end{lemma}
The operational semantics provide us the following equations to
support the proof of the above lemma.
\begin{eqnarray*}
   \close{PP}\go{a}R &=& PP\go{a}PP'~~\wedge~~R=\close{PP'}\\
    &\vee& PP\go{!}P~~\wedge~~P\go{a}P'~~\wedge~~R=P'\\
   \close{PP}\go{\omega}~0 &=& PP\go{\tick}P~~
    \wedge~~P\go{p'}0\\
    &\vee& PP\go{!}P~~\wedge~~P\go{\omega}0
\end{eqnarray*}
The block operator runs the compensation of a terminating forward
behaviour and discards the compensation of successfully completed
forward behaviour. It removes the traces of an yielding forward
behaviour.

We left two operators from the
correspondence proof presented here. First one is the choice
operator ($P~\Box Q$). The trace of choice is the union of their
traces and the operational rules shows that either process ($P$ or
$Q$) can evolve independently. Correspondence proof of this
operator is trivial. Another operator that was left is interrupt
handler ($P\rhd Q$). It is quite similar to standard sequential
composition except that the flow of control from the first to the second
process is caused by a throw ($!$) rather than a $\tick$ and
hence, showing its correspondence proof would be repetitive.

\section{Lessons Learned}\label{sec:lesson}

We have adopted a systematic approach to show the correspondence
between the two semantic models of cCSP. Traces are derived from the
operational rules and then applying induction over the traces we
showed the correspondence. Due to the way of defining operational
rules the trace derivation was done easily. We used labelled
transition system to define the operational rules. In
\cite{bruni05} operational rules are defined for a similar
language as ours but same symbol is used to define the labels of
different transition rules. However, we used special symbols for
different kinds of transitions. Transition between states are
caused by two kinds of events: normal and terminal and we used
these events as labels in our transition rules. The advantage of
this approach of defining labels is that these labels are the
traces of the transition and we can then derive these traces from
the transition rules.

The trace operators play a significant role in defining the lemmas
as well as in the correspondence proofs. The operators are used
both at the trace levels and at the process levels. All the lemmas
defined in this chapter have a common pattern applicable to both
standard and compensable processes. For example, for standard
processes $P$ and $Q$, and their traces $p$ and $q$, the lemmas
for all the operators are defined as follows:
\begin{eqnarray*}
    (P~\otimes~ Q)\go{t}0~~=~~\exists p,q\cdot t=(p\otimes
    q)~\wedge~P\go{p}0~\wedge~Q\go{q}0\\
    \mbox{(for parallel operator use~$t\in(p\otimes q)$
    instead of $t=(p\otimes q)$)}
\end{eqnarray*}

Similar definitions are also given for the forward behaviour of
compensable processes. The use of operators at both trace and
process levels allow us to apply appropriate rules for the
operators (rules for terminal and observable events from
operational and trace semantics).

The correspondence was proved by using structural induction.
First, the induction was applied on process terms of the language
and then on the derived traces. The lower level induction which is
on traces support the induction on upper level which is on process
terms

\section{Related Work}\label{sec:relwork}

The semantic correspondence presented here is based on the technique of applying structural induction. A similar approach is also applied by S. Schneider \cite{schneider:tcsp}, where an equivalence relation was established between the operational and denotational semantics of timed CSP~\cite{reed:timedcsp}\cite{timedcsp}. Operational rules are defined for timed CSP and then timed traces and refusals are extracted from the transition rules of a program, and it is shown that the pertinent information corresponds to the semantics obtained from the denotational semantic function. By applying structural induction over the terms of timed CSP, it was proved that the behaviour of the transition system is identical to those provided by the denotational semantics.

A similar problem was also investigated in \cite{tcs:metric}, where a metric structure was employed to relate the operational and denotational models of a given language. In order to relate the semantic models it was proved that the two models coincide. The denotational models were extended and structural induction was applied over the terms of the language to relate the semantic models.

Other than using induction, Hoare and He \cite{utp} presented the idea of unifying different programming paradigms and showed how to derive operational semantics from its denotational presentation of a sequential language. They derive algebraic laws from the denotational definition and then derive the operational semantics from the algebraic laws. Similar to our work, Huibiao \textit{et al.} \cite{os2ds} derived denotational semantics from operational semantics for a subset of Verilog~\cite{verilog}. However the derivation was done in a different way than our method where the authors defined transitional condition and phase semantics from the operational semantics. The denotational semantics are derived from the sequential composition of the phase semantics. The authors also derived operational semantics from denotational semantics \cite{ds2os}.

Unlike our approach, the unification between the two semantics was shown in \cite{scott-os} by extending the operational semantics to incorporate the denotational properties. The equivalence was shown for a language having simple models without any support for concurrency. Similar problem was also investigated in \cite{meyer} for a simple sequential language, which support recursion and synchronisation in the form of interleaving. The relation between operational and denotational semantics is obtained via an intermediate semantics.

\section{Concluding Remarks}\label{sec:concl}

It is of great importance to have the description of both
operational and denotational semantics. Having both of the
semantics we need to establish a relationship between these two.
Demonstrating the relationship between these two semantics of the
same language ensures the consistency of the whole  semantic
description of the language.

The main contribution of this paper is to show the correspondence
between the operational semantics and the trace semantics of a
subset of cCSP language. The correspondence is shown by deriving
the traces from the operational rules and then applying the
induction over the derived traces. Two level of induction is
applied. In one level induction is applied over the operational
rules and in the next level induction is applied over the derived
traces.

The correspondence shown here are completely done by hand which is
error prone and there are strong possibilities to miss some of the
important parts during the proof. As part of the future work our
goal is to use an automated/mechanized prover which will help us
to use the similar approach that we followed here i.e,
mathematical induction, and at the same time prove the theorems
automatically. Among several tools we are currently using
PVS (Prototype Verification System)~\cite{pvs92} for
our purpose. The specification language of PVS is based on
classical, typed, high order logic and contains the constructs
intended to ease the natural development of specification. The PVS
proof checker is interactive and provides powerful basic commands
and a mechanism for building re-usable strategies based on these.

The parallel operator of cCSP does not support synchronization on
normal events. Synchronization of events is significant for the
development of a language. Currently we are working on adding
synchronization to cCSP. Adding synchronization and then using
mechanized theorem prover for showing the correspondence will
strengthen the formal foundation of the language.




\end{document}